\documentclass[journal]{IEEEtran}
\usepackage{cite}

\ifCLASSINFOpdf 
\usepackage[pdftex]{graphicx}
\else
\usepackage[dvips]{graphicx}
\fi

\usepackage[cmex10]{amsmath}
\usepackage{algorithmic}
\usepackage[ruled,norelsize]{algorithm2e}

\usepackage{array}

\usepackage{mdwmath}
\usepackage{mdwtab}
\usepackage{todonotes}

\usepackage{eqparbox}

\ifCLASSOPTIONcompsoc
\usepackage[caption=false,font=normalsize,labelfont=sf,textfont=sf]{subfig}
\else
\usepackage[caption=false,font=footnotesize]{subfig}
\fi

\usepackage{fixltx2e}
\usepackage{url}
\usepackage{amsfonts}
\usepackage{amssymb}
\usepackage{amsthm}
\usepackage{bm}
\usepackage[ruled]{algorithm2e}
\usepackage[utf8]{inputenc}
\usepackage{booktabs}
\usepackage{url}
\usepackage{cite}
\usepackage{flushend}
\usepackage{epstopdf}

\newcounter{tempEquationCounter} 
\newcounter{thisEquationNumber}

\begin{document}
%
\title{The Missing Pillar in Quantum-Safe 6G: \\ Regulation and Global Compliance	}


\author{Adnan~Aijaz,~\IEEEmembership{Senior~Member,~IEEE}
\vspace{-0.4cm}
\thanks{The author is with the Bristol Research and Innovation Laboratory, Toshiba Europe Ltd., Bristol, BS1 4ND, UK. Contact e-mail: adnan.aijaz@toshiba-bril.com}}
\markboth{submitted for publication to IEEE}%
{Shell \MakeLowercase{\textit{et al.}}: Bare Demo of IEEEtran.cls for Journals}
%


\maketitle
\begin{abstract}
\boldmath
Sixth-generation (6G) mobile networks are expected to operate for multiple decades, supporting mission-critical and globally federated digital services. This long operational horizon coincides with rapid advances in quantum computing that threaten the cryptographic foundations of contemporary mobile systems. While post-quantum cryptography is widely recognized as a necessary technical response, its effective deployment in 6G depends equally on the evolution of regulatory policy and global compliance frameworks. This article argues that quantum-safe 6G represents a regulatory inflection point for mobile networks, as existing compliance models shaped by static cryptographic assumptions, incremental evolution, and point-in-time certification are poorly suited to long-term quantum risk. Building on an analysis of baseline telecom compliance challenges, the evolution of security regulation from 2G to 5G, and the regulatory impact of post-quantum cryptography adoption, the article shows why incremental regulatory extensions are insufficient. To address this gap, the article advances a compliance-by-design perspective in which regulatory requirements are treated as system-level design constraints, emphasizing cryptographic agility, lifecycle-aware governance, continuous compliance observability, and interoperability-driven global assurance, and concludes by examining the risks of fragmented global compliance for quantum-safe 6G networks.

\end{abstract}



%
\IEEEpeerreviewmaketitle

\section{Introduction}
\IEEEPARstart{6}{G} networks are expected to operate for multiple decades, supporting mission-critical services, massive machine-type communications, and globally federated digital infrastructures~\cite{ieee6g}. This extended operational horizon coincides with rapid advances in quantum computing that threaten the cryptographic foundations of contemporary mobile systems. Quantum algorithms, most notably Shor’s algorithm, are widely expected to render deployed public-key cryptosystems such as RSA and elliptic curve cryptography (ECC) insecure within the lifetime of future mobile networks~\cite{shor1997,nistPQC}. Unlike previous security upgrades, the quantum threat undermines foundational trust mechanisms rather than isolated protocol vulnerabilities.

As a result, the effectiveness of quantum-safe 6G depends not only on the adoption of post-quantum cryptographic primitives, but also on the capacity of regulatory and compliance frameworks to anticipate, enforce, and adapt to long-term security risk. Historically, mobile security regulation has prioritized interoperability, backward compatibility, and incremental evolution. While these principles enabled rapid global adoption of earlier generations, they are poorly matched to the disruptive, time-sensitive, and systemic nature of quantum threats.

Recent national and international policy developments underscore the urgency of this challenge. Governments and cybersecurity authorities have begun publishing structured timelines for migration to post-quantum cryptography, recognizing that transition will span many years and require coordinated action across vendors, operators, standards bodies, and regulators~\cite{ncscPQC,nistTransition}. At the same time, the telecommunications sector already operates under significant regulatory pressure related to data protection, supply-chain security, geopolitical constraints, sustainability, and operational resilience~\cite{gdpr,ccpa}. Quantum-safe 6G will therefore emerge not in a regulatory vacuum, but within an already fragmented and compliance-constrained ecosystem.

This article argues that quantum-safe 6G represents a regulatory inflection point for mobile networks. Rather than treating compliance as a post-deployment obligation, it examines how regulatory policy and global compliance must be integrated as system-level design considerations. By framing regulation as an architectural constraint and enabler, the article advances a compliance-by-design perspective as a prerequisite for sustainable, globally interoperable quantum-safe 6G systems.
The main contributions of this article are as follows:
\begin{itemize}
    \item It positions regulatory policy and global compliance as key design dimensions for quantum-safe 6G, extending algorithm-centric security research.
    \item It examines how post-quantum cryptography and long-lived confidentiality requirements amplify existing telecom compliance challenges.
    \item It defines core principles for compliance-by-design, including cryptographic agility, lifecycle-aware governance, and continuous compliance observability.
    \item It presents a lightweight evaluation framework with indicative results demonstrating improved adaptability and reduced compliance risk.
\end{itemize}

The article begins by outlining existing regulatory and compliance challenges in the telecommunications industry and their evolution across mobile generations. It then explores the regulatory implications of post-quantum cryptography and introduces compliance-by-design as a guiding principle for achieving resilient and globally interoperable quantum-safe 6G networks.

\section{Baseline Compliance Challenges in the Telecom Industry}

Even prior to the emergence of quantum threats, the telecommunications industry operates under a dense, fragmented, and continuously evolving regulatory landscape. Telecom operators manage vast volumes of sensitive customer data, including location information, communication metadata, and personally identifiable information, often across national borders and legal jurisdictions. Compliance with data protection regimes such as the EU General Data Protection Regulation (GDPR) and the California Consumer Privacy Act (CCPA) imposes stringent requirements on data handling, retention, breach notification, and user consent, frequently with differing interpretations and enforcement practices across regions~\cite{gdpr,ccpa}. Ensuring consistent compliance across heterogeneous networks and multinational operations already represents a significant operational burden.

Regulatory fragmentation further complicates compliance obligations. Global telecom operators must navigate a complex mix of national and regional frameworks governing cybersecurity, lawful access, spectrum allocation, infrastructure security, and consumer protection. Cross-border operations, including roaming, interconnection, and federated service delivery, introduce additional layers of compliance overhead, as operators must simultaneously satisfy home- and host-country requirements that may not be fully aligned~\cite{ituRegulatory}. In practice, this fragmentation often leads to duplicated controls, increased audit complexity, and conservative design choices that prioritize regulatory risk mitigation over technical optimization.

Geopolitical tensions and economic sanctions introduce another dimension of compliance complexity. Export controls, investment restrictions, and vendor-specific bans can shift rapidly in response to international developments, forcing operators to reassess supply-chain relationships and network dependencies on short notice. Given the telecommunications industry’s reliance on a relatively small number of vendors for critical infrastructure, ensuring ongoing compliance with such measures is particularly challenging~\cite{enisaSupplyChain}. These constraints not only affect procurement decisions but also influence long-term network evolution, upgrade cycles, and architectural diversity.

Additional regulatory pressures arise from emerging governance domains. The increasing adoption of artificial intelligence and machine learning for network optimization, fault management, and customer interaction introduces new regulatory expectations related to transparency, fairness, explainability, and accountability. At the same time, environmental, social, and governance (ESG) requirements place growing emphasis on sustainability, energy efficiency, and climate-related risk disclosure. As telecom networks are widely classified as critical infrastructure, operators are also subject to stringent resilience, availability, and incident-reporting obligations, further increasing compliance scope and scrutiny.

Taken together, these baseline challenges demonstrate that telecom operators are already operating near the practical limits of regulatory complexity, balancing technical innovation against overlapping and sometimes conflicting compliance demands. Quantum-safe 6G will not replace these obligations but will compound them by introducing long-term cryptographic risk, new assurance requirements, and additional coordination challenges across standards bodies, regulators, and vendors. This cumulative burden, summarized across mobile generations in Table~\ref{tab:evolution_security_regulation}, underscores the need to reassess how compliance is integrated into the design and governance of future mobile systems.

\begin{table*}[t]
\caption{Evolution of Security Regulation and Compliance Models in Mobile Networks}
\label{tab:evolution_security_regulation}
\centering
\renewcommand{\arraystretch}{1.2}
\begin{tabular}{|c|p{3.5cm}|p{4cm}|p{4.5cm}|}
\hline
\textbf{Generation} & \textbf{Primary Regulatory Focus} & \textbf{Dominant Compliance Model} & \textbf{Structural Limitation} \\
\hline
2G & Service availability and lawful access & Minimal assurance; operator trust assumed & Weak cryptography; reactive mitigation; no formal assurance \\
\hline
3G & Authentication and interoperability & Standards-driven compliance & Limited cryptographic agility; static threat assumptions \\
\hline
4G & Assurance and certification & Point-in-time certification & Slow cryptographic transitions; lifecycle blind spots \\
\hline
5G & System-level risk management & Continuous assurance (classical) & Anchored in classical cryptography; static PKI assumptions \\
\hline
6G & Long-term resilience and adaptability & Lifecycle-aware, compliance-by-design & Regulatory harmonization and observability challenges \\
\hline
\end{tabular}
\end{table*}

\section{Evolution of Security Regulation in Mobile Networks}

Security regulation in mobile networks has evolved incrementally from 2G to 5G, closely tracking the prevailing threat models, deployment realities, and economic priorities of each generation. Throughout this evolution, regulatory frameworks have largely assumed that cryptographic foundations would remain stable over the lifetime of a generation, an assumption that becomes increasingly problematic in the face of quantum threats.

\subsection{Early Generations: Availability, Interoperability, and Implicit Trust}

In 2G and early 3G systems, regulatory attention focused primarily on service availability, lawful access, and global interoperability. Security mechanisms were largely standardized through technical specifications, with limited regulatory scrutiny of cryptographic robustness or adversarial capabilities~\cite{3gppSecurityOverview}. Cryptography was treated as an enabling feature rather than a compliance-critical control, and trust assumptions, particularly regarding network operators and infrastructure vendors, were largely implicit.

When vulnerabilities emerged, responses were typically reactive and localized. Operators deployed mitigations through configuration changes or proprietary extensions, and regulators rarely mandated coordinated, system-wide security upgrades. As a result, security regulation in early generations was characterized by minimal assurance requirements and a strong reliance on backward compatibility.

\subsection{4G: Standardization, Certification, and Static Assurance}

The transition to 4G marked a shift toward more formalized security regulation. Regulators increasingly relied on international standards bodies, such as 3GPP and ETSI, to define security baselines that could be uniformly implemented across markets~\cite{3gppLTE}. Certification and assurance frameworks emerged, aiming to improve consistency and interoperability while reducing operator-specific deviations.

However, these frameworks remained largely static and assumption-driven. Compliance was typically demonstrated through point-in-time certification against approved algorithms and protocol profiles. Cryptographic transitions were infrequent, slow to propagate, and often driven by standards updates rather than evolving threat intelligence. Long-term adversarial models, including threats capable of retroactively compromising encrypted data, were generally outside the regulatory scope.

\subsection{5G: System-Level Risk Management and Expanded Regulatory Scope}

5G represents a significant maturation of security regulation, reflecting growing recognition of mobile networks as critical national infrastructure. Regulatory frameworks expanded beyond protocol security to encompass system-level risk management, supply-chain security, virtualization, and continuous assurance mechanisms~\cite{3gpp33501,ituIMT2020}. Concepts such as security-by-design, network slicing isolation, and zero-trust principles gained prominence in both standards and regulatory guidance.

Despite these advances, 5G security regulation remains fundamentally anchored in classical computational assumptions. Public-key infrastructures, key agreement protocols, and long-term confidentiality guarantees continue to rely on cryptographic primitives vulnerable to future quantum adversaries. Moreover, certification and compliance processes still emphasize static validation at deployment, with limited mechanisms to mandate or verify cryptographic agility over the network lifecycle.

\subsection{Structural Limitations of Legacy Regulatory Models}

Across generations, several structural characteristics persist in mobile security regulation. Regulatory processes tend to lag behind technological change, favor backward compatibility to preserve global interoperability, and rely on static certification models that assume cryptographic stability. While these characteristics supported the rapid global adoption of earlier generations, they introduce systemic risk in the context of quantum threats.

As summarized in Table~\ref{tab:evolution_security_regulation}, existing regulatory approaches are ill-suited to address adversaries capable of undermining foundational cryptographic assumptions across decades of deployed infrastructure. These limitations, regulatory latency, backward, compatibility bias, and static assurance—become critical challenges for quantum-safe 6G, motivating the need for fundamentally different compliance and governance models.

\section{Quantum-Safe Cryptography as a Regulatory Disruptor}

The introduction of post-quantum cryptography (PQC) represents a structural disruption to existing telecommunications compliance models rather than a routine security upgrade. Unlike previous cryptographic transitions, which were often driven by incremental improvements or localized vulnerabilities, the quantum threat challenges the long-term validity of foundational public-key mechanisms. As a result, PQC adoption forces regulators and operators to reconsider how cryptographic compliance is defined, validated, and enforced over the lifecycle of mobile networks.

\subsection{PQC Migration as a Lifecycle Compliance Challenge}

PQC migration is not a one-time upgrade but a multi-phase process involving cryptographic inventory, hybrid operation, algorithm replacement, and eventual deprecation of legacy mechanisms~\cite{nistTransition}. Each phase introduces distinct compliance obligations, ranging from documenting cryptographic dependencies to demonstrating readiness for future algorithm transitions. National guidance increasingly reflects this long-term perspective, emphasizing early preparation, staged migration, and explicit planning horizons extending into the 2030s~\cite{ncscPQC}.

From a regulatory standpoint, this lifecycle view departs from traditional compliance models that focus on point-in-time certification. Operators are increasingly expected to show not only that approved algorithms are deployed, but also that systems can evolve securely as cryptographic standards change. This shift places new emphasis on cryptographic agility as a compliance requirement rather than an optional engineering feature.

\subsection{Performance and Interoperability as Compliance Constraints}

Beyond migration planning, PQC introduces performance and interoperability challenges that directly affect regulatory compliance. Many PQC algorithms rely on larger key sizes, increased message exchanges, and higher computational complexity than their classical counterparts. These characteristics can impact latency-sensitive control-plane procedures, increase signaling overhead, and strain resource-constrained devices at the network edge~\cite{etsiPQC}.

From a compliance perspective, such impacts translate into risks where service-level agreements, energy efficiency targets, or certification thresholds are exceeded. Regulatory frameworks often assume stable cryptographic overheads, an assumption that no longer holds in a quantum-safe context. As summarized in Table~\ref{tab:compliance_before_after}, these deviations challenge existing compliance baselines and require regulators to reconcile security objectives with operational performance constraints.

\subsection{Hybrid Cryptography and Transitional Assurance Complexity}

To address the risk of harvest-now-decrypt-later attacks, hybrid cryptographic deployments that combine classical and post-quantum mechanisms are widely recommended during transition periods~\cite{moscaHNDL}. While hybrid approaches offer risk mitigation, they introduce additional complexity for compliance validation. Regulators must assess the combined assurance properties of multiple cryptographic mechanisms and determine whether security guarantees are additive, redundant, or mutually dependent.

This raises non-trivial questions for compliance enforcement. For example, it is unclear how long legacy algorithms may remain acceptable within hybrid configurations, or what criteria should trigger mandatory deprecation. Without clear regulatory guidance, hybrid deployments risk becoming prolonged transitional states, increasing operational complexity and compliance uncertainty.

\subsection{Vendor Readiness and Supply-Chain Compliance Risk}

Vendor readiness further complicates quantum-safe compliance. Telecom infrastructure frequently depends on hardware-embedded cryptographic functions with long deployment and upgrade cycles. In many cases, cryptographic capabilities are constrained by silicon design choices made years in advance. Misalignment between regulatory timelines, standards evolution, and vendor roadmaps can therefore leave operators unable to meet emerging compliance expectations despite good-faith efforts.

From a regulatory perspective, this creates a new category of compliance risk rooted in supply-chain dependencies rather than operator negligence. Addressing this risk may require regulators to place greater emphasis on vendor transparency, forward-compatibility guarantees, and evidence of long-term cryptographic support as part of compliance assessments.

\subsection{Implications for Regulatory Models}

Taken together, these factors illustrate why PQC acts as a regulatory disruptor for telecom networks. Existing compliance frameworks, designed around static algorithms, predictable performance characteristics, and infrequent cryptographic change, are poorly suited to the dynamic and long-term nature of quantum-safe security. Without adaptation, these frameworks risk either inhibiting timely PQC adoption or allowing prolonged exposure to quantum-era threats.

This disruption underscores the need for regulatory models that are explicitly lifecycle-aware, performance-conscious, and supply-chain-informed, setting the stage for compliance-by-design approaches in quantum-safe 6G systems.

\begin{table*}[t]
\caption{Compliance Implications of Post-Quantum Cryptography in Telecom Networks}
\label{tab:compliance_before_after}
\centering
\renewcommand{\arraystretch}{1.2}
\begin{tabular}{|p{3.5cm}|p{5.5cm}|p{5.5cm}|}
\hline
\textbf{Compliance Dimension} & \textbf{Pre-Quantum Compliance Assumptions} & \textbf{Quantum-Safe Compliance Implications} \\
\hline
Cryptographic stability & Long-lived algorithms; rare transitions & Frequent transitions; mandatory cryptographic agility \\
\hline
Performance baselines & Stable latency, compute, and energy overheads & Increased and variable overheads affecting SLAs and certification \\
\hline
Assurance validation & Static certification and periodic audits & Continuous, evidence-driven compliance verification \\
\hline
Hybrid operation & Limited or temporary coexistence of mechanisms & Prolonged hybrid phases requiring explicit regulatory control \\
\hline
Supply-chain assurance & Vendor trust and incremental upgrades & Dependency on vendor PQC readiness and forward-compatibility \\
\hline
Risk horizon & Forward-looking confidentiality only & Harvest-now–decrypt-later and retroactive exposure \\
\hline
\end{tabular}
\end{table*}

\begin{figure}
\centering
\includegraphics[width=1\linewidth]{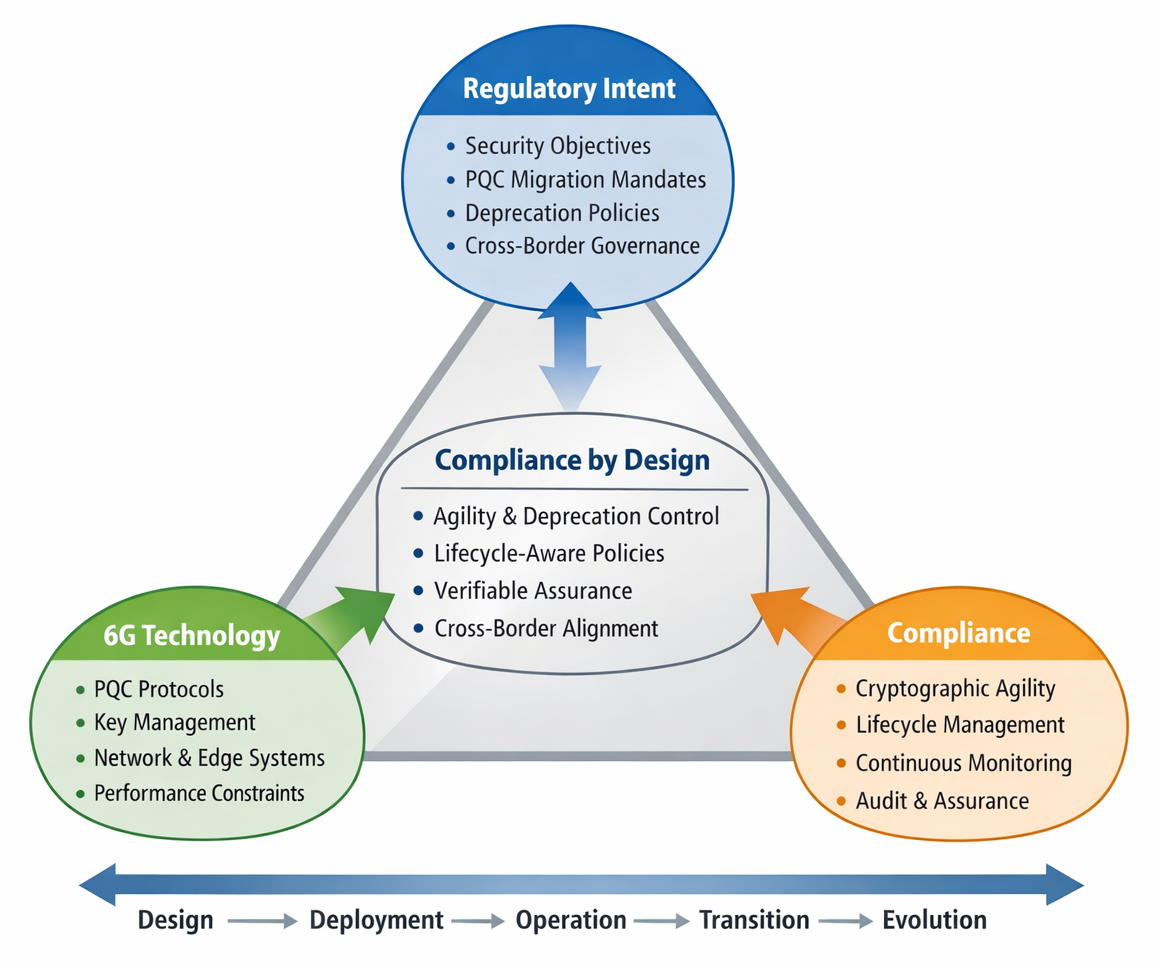}
    \caption{Compliance-by-design for quantum-safe 6G with lifecycle overlay.}
    \label{comp}
\end{figure}

\section{Compliance-by-Design for Quantum-Safe 6G}

The cumulative analysis of baseline telecom compliance challenges, the historical evolution of security regulation, and the disruptive impact of post-quantum cryptography demonstrates that incremental regulatory extensions are insufficient for quantum-safe 6G. Instead, compliance must be treated as an explicit design objective, co-equal with performance, scalability, and reliability, and addressed continuously throughout the system lifecycle. This compliance-by-design approach requires rethinking how regulatory intent is translated into architectural features, operational processes, and assurance mechanisms.

At a conceptual level, compliance-by-design can be understood as an integrative function that mediates between regulatory policy and 6G system architecture across time, as illustrated in Fig.~\ref{comp}. Regulatory frameworks define long-term security objectives, governance constraints, and migration timelines, while 6G technologies implement cryptographic, networking, and operational capabilities under strict performance and scalability constraints. Compliance-by-design bridges these domains by embedding lifecycle-aware, observable, and enforceable assurance into system design and operation, ensuring that regulatory intent remains actionable as networks evolve.

\subsection{Embedding Regulation into System Design}

In previous mobile generations, regulatory compliance was typically addressed after system design through certification, audits, and operator-specific controls. Quantum-safe 6G inverts this relationship. Long-term cryptographic risk, regulatory timelines for post-quantum migration, and cross-border interoperability requirements impose constraints that must shape protocol design, key management architectures, and system modularity from the outset.

Treating regulation as a first-class system constraint implies that compliance objectives such as cryptographic agility, algorithm deprecation capability, and assurance transparency are embedded directly into standards and reference architectures. This shift mirrors how availability and safety constraints are treated in other critical infrastructures and reflects the reality that non-compliant systems may be operationally unusable in regulated environments.

\subsection{Cryptographic Agility as a Compliance Metric}

Cryptographic agility is a cornerstone of compliance-by-design for quantum-safe 6G. While agility is often discussed qualitatively, it can also be framed as a compliance performance metric. Relevant dimensions include the time required to replace cryptographic algorithms, the scope of systems affected by a transition, and the operational impact on service continuity.

From a regulatory perspective, these dimensions translate into measurable expectations such as acceptable transition windows, limits on prolonged hybrid operation, and requirements for maintaining accurate cryptographic inventories. Systems that require hardware replacement or extended service disruption to achieve compliance introduce regulatory risk regardless of their nominal cryptographic strength. In this sense, agility becomes a proxy for regulatory resilience.

\subsection{Lifecycle-Aware and Context-Aware Compliance}

Quantum-safe compliance cannot be static. A lifecycle-aware approach explicitly distinguishes between pre-deployment assurance, transitional hybrid operation, quantum-safe baseline enforcement, and post-deployment adaptation, with each phase carrying distinct compliance obligations and risk profiles. Regulators may tolerate hybrid cryptography during transition periods, but only when accompanied by demonstrable plans, milestones, and deprecation mechanisms.

In 6G, lifecycle awareness must also be context-aware. Network slicing, edge computing, and federated service delivery imply that different parts of the network may operate under different regulatory regimes simultaneously. Compliance mechanisms must therefore support slice-specific assurance, jurisdiction-aware policy enforcement, and cross-domain accountability without undermining interoperability or performance.

\subsection{Compliance Observability and Evidence Generation}

A defining feature of compliance-by-design is observability. Traditional compliance models rely heavily on periodic audits and static documentation, which are poorly suited to long-lived systems facing evolving cryptographic threats. Quantum-safe 6G demands continuous, machine-verifiable evidence of compliance-relevant system state.

Such evidence may include cryptographic inventories, algorithm usage telemetry, key lifecycle status, and policy enforcement logs. From a qualitative perspective, the timeliness, granularity, and trustworthiness of this evidence determine how quickly regulators and operators can detect non-compliance or emerging risk. Quantitatively, observability can be assessed through metrics such as detection latency, coverage across network components, and assurance update frequency.

\subsection{Balancing Security, Performance, and Compliance}

Compliance-by-design also requires explicit trade-off management between security objectives, network performance, and regulatory obligations. Post-quantum cryptographic overheads, hybrid configurations, and continuous assurance mechanisms consume computational, bandwidth, and operational resources. Rather than treating these costs as externalities, compliance-aware design incorporates them into performance planning and capacity models.

This perspective enables regulators and operators to reason about acceptable trade-offs, such as temporary performance degradation during cryptographic transitions or differentiated compliance profiles for latency-critical versus delay-tolerant services. It shifts compliance discussions away from binary pass–fail assessments toward risk-informed, performance-aware governance.

\subsection{Implications for Standards and Certification}

Adopting compliance-by-design has significant implications for standards bodies and certification schemes. Standards must evolve to specify not only approved algorithms but also required agility mechanisms, observability hooks, and lifecycle management capabilities. Certification, in turn, must move beyond one-time approval toward continuous or periodic re-validation aligned with evolving threat and policy landscapes.

These changes align regulatory practice with the realities of quantum-safe 6G, where long-term security assurance depends less on static correctness and more on sustained adaptability, transparency, and accountability.

\begin{table*}[t]
\caption{Evolutionary and Revolutionary Regulatory Approaches for Quantum-Safe 6G}
\label{tab:regulatory_approaches}
\centering
\renewcommand{\arraystretch}{1.2}
\begin{tabular}{|p{3.5cm}|p{5.5cm}|p{5.5cm}|}
\hline
\textbf{Aspect} & \textbf{Evolutionary Approach} & \textbf{Revolutionary Approach} \\
\hline
Regulatory philosophy & Extend 5G compliance frameworks & Redefine compliance objectives for long-term quantum resilience \\
\hline
Cryptographic governance & Algorithm-centric approval & Lifecycle-centric agility and deprecation management \\
\hline
Compliance validation & Static certification with updates & Continuous, observability-driven assurance \\
\hline
Performance considerations & Implicit, secondary to security & Explicit trade-offs between security, performance, and compliance \\
\hline
Deployment risk & Lower short-term disruption & Higher initial complexity; lower long-term risk \\
\hline
Global alignment & Easier short-term harmonization & Requires early cross-border coordination and mutual recognition \\
\hline
\end{tabular}
\end{table*}

\section{Discussion and Key Insights}

Quantum-safe 6G exposes a fundamental gap between traditional regulatory models and the realities of long-lived, globally interconnected, and software-defined mobile networks. Existing compliance frameworks were largely designed for environments in which cryptographic assumptions remained stable over the operational lifetime of a generation. The prospect of practical quantum adversaries invalidates this premise, requiring regulation to contend with long-term confidentiality risks, inevitable cryptographic change, and prolonged transition periods.

Addressing this gap demands closer and more structured alignment between regulators, standards bodies, and industry stakeholders. Standards alone cannot guarantee compliance, while regulation that is decoupled from technical realities risks being either ineffective or disruptive. Quantum-safe 6G therefore necessitates a tighter coupling between policy objectives and system design, where regulatory intent is translated into enforceable architectural and operational properties.

Several open challenges emerge from this analysis. One is the need to define machine-verifiable compliance evidence that can support continuous assurance at scale. Periodic audits and static documentation are poorly suited to environments characterized by cryptographic agility, network slicing, and dynamic service instantiation. Another challenge lies in harmonizing cross-border assurance expectations without requiring full legal convergence. As demonstrated by existing telecom practices such as roaming, interoperability-driven governance may offer a more scalable path to global compliance than purely legal harmonization.

Managing cryptographic deprecation is a further unresolved issue. Quantum-safe 6G systems must support orderly, regulator-visible deprecation of vulnerable algorithms without destabilizing critical services or creating prolonged hybrid states that increase complexity and risk. Achieving this balance will require regulatory mechanisms that explicitly address transition timelines, performance impacts, and ecosystem dependencies.

Overall, the analysis suggests that quantum-safe 6G represents not only a security transition but also a regulatory inflection point. The success of future mobile systems will depend on whether compliance frameworks can evolve from static, deployment-centric models toward adaptive, lifecycle-aware governance that matches the temporal and systemic nature of quantum risk.

The following observations summarize the key insights of this work:

\begin{itemize}
    \item Quantum-safe 6G challenges the assumptions underpinning existing regulatory and compliance models, particularly the expectation of long-term cryptographic stability.
    \item Effective global compliance in telecommunications is more likely to emerge from interoperability-driven governance mechanisms than from full legal harmonization.
    \item Continuous, machine-verifiable compliance evidence is essential for managing long-lived quantum risk in dynamic 6G environments.
    \item Cryptographic deprecation and hybrid operation introduce regulatory challenges that extend beyond technical implementation and require explicit governance.
    \item Compliance-by-design and lifecycle-aware regulation are necessary to avoid accumulating security and compliance debt in future mobile networks.
\end{itemize}

\subsection{Risks of Fragmented Global Compliance}
If global compliance for quantum-safe 6G is not achieved, the most immediate risk is the emergence of fragmented security baselines across jurisdictions and operators. In such a scenario, cryptographic transitions would occur at uneven rates, leading to prolonged hybrid deployments and inconsistent assurance levels across roaming, interconnection, and federated service environments. From an operational perspective, this fragmentation increases attack surface and complicates threat modeling, as adversaries can target the weakest compliant domains to compromise cross-border services. Even without large-scale quantum computers, the persistence of quantum-vulnerable public-key mechanisms enables harvest-now–decrypt-later attacks, exposing long-lived data such as identity credentials, location histories, and signaling metadata to future retrospective compromise.

The longer-term impact of non-aligned compliance is systemic and economic. Telecom networks are critical infrastructure, and loss of trust in their long-term confidentiality and integrity has cascading effects on dependent sectors such as energy, transportation, healthcare, and financial services. Fragmented compliance regimes would also increase operational costs, as operators are forced to maintain region-specific cryptographic profiles, certification processes, and assurance mechanisms. Empirically, compliance-driven network adaptations already account for a significant portion of operational expenditure in globally deployed networks; introducing divergent quantum-safety requirements would further increase this burden while reducing interoperability and innovation velocity. In the absence of harmonized, lifecycle-aware compliance frameworks, the risk is not only delayed quantum-safe adoption but the accumulation of long-term security and compliance debt that becomes progressively harder—and more expensive—to unwind as 6G systems mature.

\section{Evaluation of Compliance-by-Design}

\begin{figure}[t]
\centering
\includegraphics[width=1\linewidth]{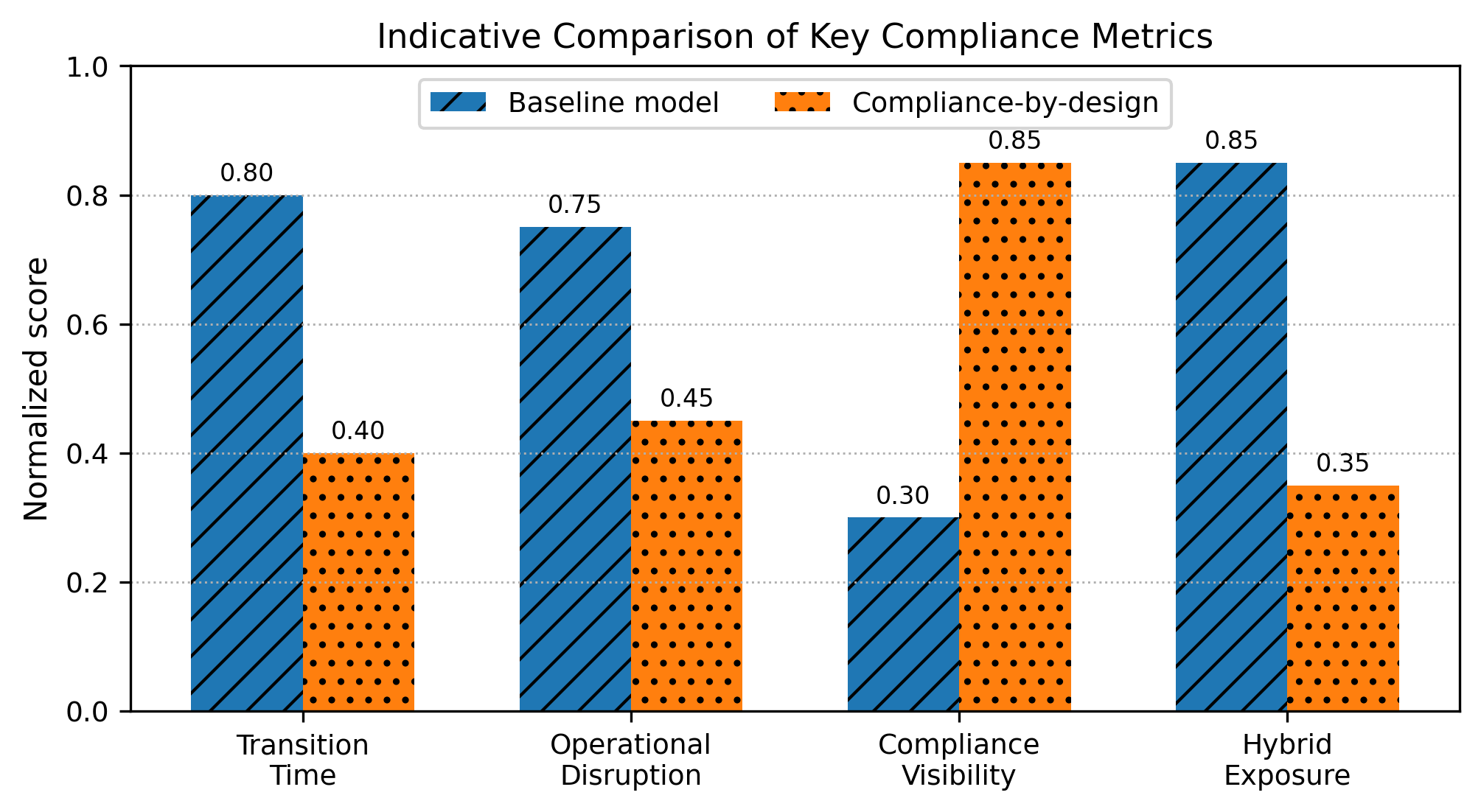}
\caption{Indicative comparison of key compliance metrics for baseline and compliance-by-design models.}
\label{fig:eval_metrics}
\end{figure}

\begin{figure}[t]
\centering
\includegraphics[width=1\linewidth]{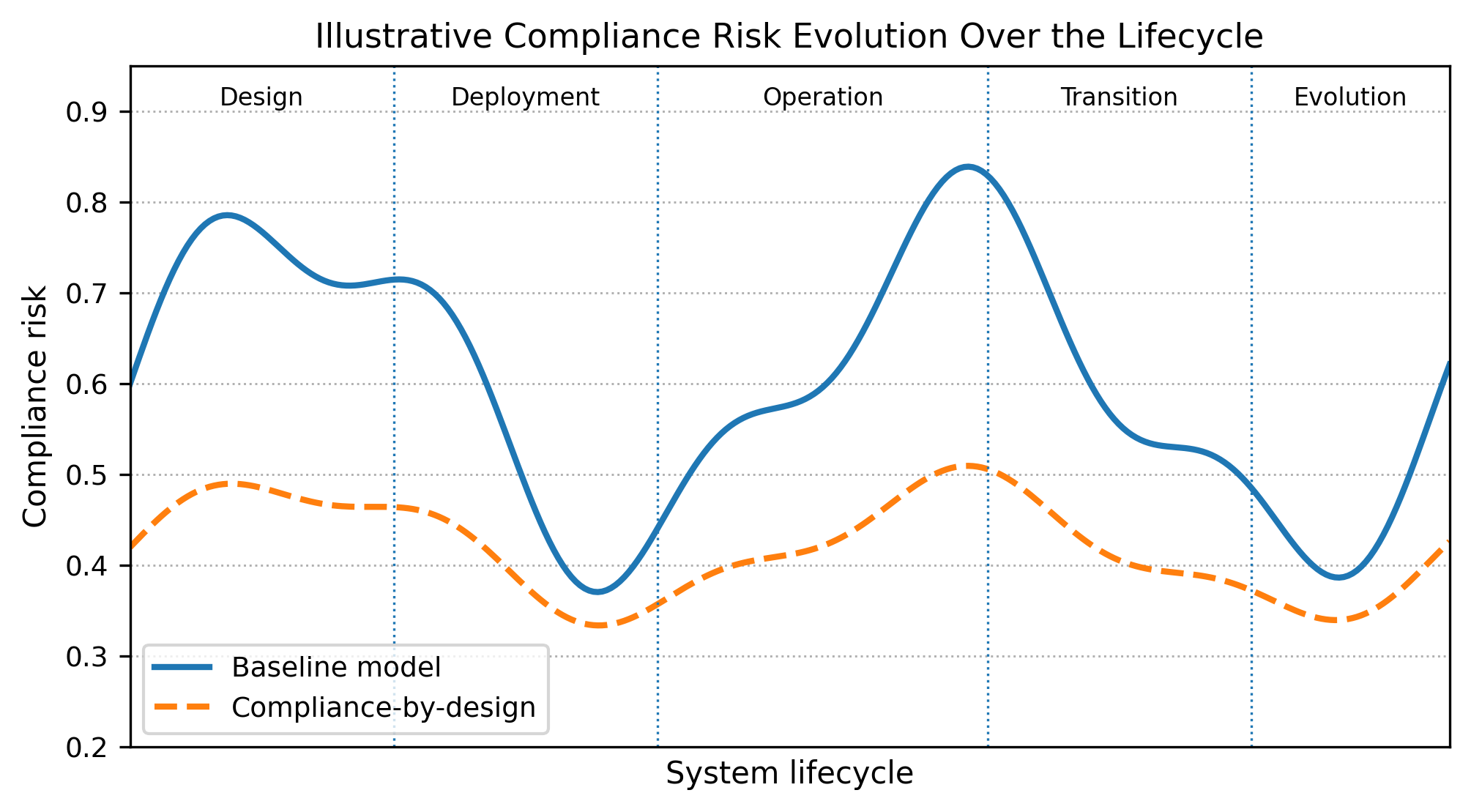}
\caption{Illustrative compliance risk evolution over the system lifecycle.}
\label{fig:eval_risk}
\end{figure}

To assess the effectiveness of compliance-by-design, we adopt a lightweight evaluation framework based on lifecycle-aware metrics that capture system adaptability, operational impact, and assurance visibility under a representative post-quantum migration scenario. Specifically, we consider cryptographic transition time, operational disruption, compliance visibility, and hybrid exposure duration as key indicators of regulatory readiness. Fig.~\ref{fig:eval_metrics} presents an indicative comparison between traditional compliance models and compliance-by-design, showing reduced transition time and disruption, along with improved observability and controlled hybrid operation. Complementarily, Fig.~\ref{fig:eval_risk} illustrates the evolution of compliance risk over time, where compliance-by-design enables smoother, more predictable risk profiles across lifecycle phases. While these results are illustrative, they highlight that embedding compliance into system design can measurably improve adaptability and reduce long-term regulatory and security exposure.

\section{Concluding Remarks}

Quantum-safe 6G cannot be achieved through cryptographic innovation alone. The transition to post-quantum security exposes fundamental limitations in existing regulatory and compliance frameworks, which were designed for static cryptographic assumptions and shorter system lifecycles. As mobile networks evolve into long-lived, globally federated digital infrastructure, these limitations become a source of systemic risk.

This work argues that regulatory policy and global compliance must evolve toward adaptive, lifecycle-aware, and system-integrated models that explicitly account for long-term cryptographic uncertainty. Treating regulation as a design dimension, rather than a post-deployment constraint, enables closer alignment between standards, governance, and operational realities. Such an approach supports orderly cryptographic transition, improves transparency and accountability, and reduces the accumulation of long-term security and compliance debt. Indicative evaluation results further highlight the potential of compliance-by-design to improve adaptability and reduce compliance risk.

Ultimately, the sustainability and trustworthiness of quantum-safe 6G will depend on whether regulatory and compliance frameworks evolve at a pace and scale commensurate with the technologies they govern. Achieving this alignment is essential for managing quantum-era risk while preserving global interoperability and trust in future mobile networks.

\bibliographystyle{IEEEtran}

\bibliography{IEEEabrv,QS6G}
\end{document}